\documentclass[12pt]{article}
\usepackage{amssymb}
\usepackage{amsmath}

\usepackage[normalem]{ulem}

\numberwithin{equation}{section}
\newcommand{\be}{\begin{eqnarray}}
\newcommand{\ee}{\end{eqnarray}}
\newcommand{\non}{\nonumber}
\newcommand{\id}{\mathbb{I}}

\newcommand{\diag}{\mathop{\rm diag}\nolimits}
\newcommand{\am}{\mathop{\rm am}\nolimits}
\newcommand{\cn}{\mathop{\rm cn}\nolimits}
\newcommand{\sn}{\mathop{\rm sn}\nolimits}
\newcommand{\dn}{\mathop{\rm dn}\nolimits}

\newcommand{\ud}{\langle 12\rangle}

\newcommand{\mF}{\mathcal{F}}
\newcommand{\mE}{\mathcal{E}}
\newcommand{\mH}{\mathcal{H}}
\newcommand{\mP}{\mathcal{P}}

\usepackage[usenames,dvipsnames]{color}

\newcommand\blfootnote[1]{%
  \begingroup
  \renewcommand\thefootnote{}\footnote{#1}%
  \addtocounter{footnote}{-1}%
  \endgroup
}

\begin{document}

\begin{titlepage}
\strut\hfill UMTG--285
\vspace{.5in}
\begin{center}

\LARGE 
Fusion for AdS/CFT boundary S-matrices\\
\vspace{1in}
\large 
Rafael I. Nepomechie \footnote{
Physics Department,
P.O. Box 248046, University of Miami, Coral Gables, FL 33124 USA}
and Rodrigo A. Pimenta ${}^{1,}$\footnote{
Departamento de F\'{i}sica, Universidade Federal de S\~{a}o Carlos, Caixa Postal 676, 
CEP 13569-905, S\~{a}o Carlos, Brasil}\\[0.8in]
\end{center}

\vspace{.5in}

\begin{abstract}
We propose a fusion formula for AdS/CFT worldsheet boundary S-matrices. We show 
that, starting from the fundamental $Y=0$ boundary S-matrix, this 
formula correctly reproduces the two-particle bound-state boundary 
S-matrices.
\end{abstract}

\blfootnote{e-mail addresses: {\tt nepomechie@physics.miami.edu, pimenta@df.ufscar.br}}

\end{titlepage}

\setcounter{footnote}{0}

\section{Introduction}

The formation of bound states (``fusion'') is a ubiquitous phenomenon
in quantum field theory.  If the theory is integrable
\cite{Zamolodchikov:1978xm}, then the factorized bulk S-matrices of
the bound-state particles can be determined in terms of the
corresponding S-matrices of the fundamental particles
\cite{Karowski:1978ps}.  This phenomenon was abstracted in
\cite{Kulish:1981gi, Kulish:1981bi} into a general ``fusion
procedure'' for constructing higher-dimensional R-matrices (solutions
of the Yang-Baxter equation) starting from a fundamental R-matrix.
For boundary S-matrices/K-matrices, i.e. solutions of the boundary
Yang-Baxter equation \cite{Cherednik:1985vs, Sklyanin:1988yz,
Ghoshal:1993tm}, an analogous fusion procedure was formulated in
\cite{Mezincescu:1990fc, Mezincescu:1991ke, Zhou:1995zy}.

In order to carry out the fusion procedure \cite{Kulish:1981gi}, the fundamental
R-matrix should satisfy a certain technical requirement: namely, it must
degenerate into a projection operator for some value(s) of the
spectral parameter. Many R-matrices fulfill this requirement;
and this fusion procedure has proved to be very useful: it not only
generates new solutions of the Yang-Baxter equation, but it also leads
to a hierarchy of commuting transfer matrices that can be used to solve
the corresponding integrable models (see e.g. \cite{Kuniba:2010ir}).

However, the AdS/CFT worldsheet bulk S-matrix \cite{Staudacher:2004tk,
Beisert:2005tm}, which plays a key role in the understanding of
integrability in AdS/CFT \cite{Beisert:2010jr}, does not satisfy this
requirement.  This apparent failure of the fusion procedure has been
quite puzzling, since bound states do form in this model
\cite{Dorey:2006dq, Arutyunov:2007tc}, and their bulk S-matrices have
been determined \cite{Arutyunov:2008zt, Arutyunov:2009mi}, albeit by
other means.

This puzzle was recently resolved by Beisert, de Leeuw and Nag
\cite{Beisert:2015msa}, who showed that one can relax the requirement
that the R-matrix degenerates into a projector.  In particular, they
proposed a new bulk fusion formula, which generates a bound-state
AdS/CFT bulk S-matrix \cite{Arutyunov:2008zt} from the fundamental
one.  As a bonus, the resulting fused matrix automatically has the
correct dimensions - the additional similarity transformation and
subsequent elimination of null rows and columns that are implicit in
the original approach \cite{Kulish:1981gi} are not needed. (A similar 
fusion formula was proposed for the XXX R-matrix in \cite{Gohmann:2010se}.)

Factorized {\em boundary} S-matrices also play an interesting role in
AdS/CFT (see e.g. \cite{Hofman:2007xp, Zoubos:2010kh}); and AdS/CFT
boundary S-matrices for bound states have also been determined 
\cite{Correa:2009mz, Ahn:2010xa, Palla:2011eu}.  (See also 
\cite{MacKay:2010ey, MacKay:2011av, Correa:2011nz, MacKay:2011zs, 
Regelskis:2011fa, Gomez:2014uxa} and references therein.)
The main purpose of this note is to
propose a new fusion formula for boundary S-matrices, which generates
two-particle bound-state AdS/CFT boundary S-matrices 
from the fundamental one.

The outline of this paper is as follows.  In Section \ref{sec:bulk} we
briefly review the new bulk fusion procedure formulated in
\cite{Beisert:2015msa}.  However, we work with different conventions,
which we find more convenient.  In Section \ref{sec:boundary} we
present the corresponding boundary fusion formula, whose proof is
relegated to an appendix.  We then show that, starting from the
fundamental $Y=0$ boundary S-matrix \cite{Hofman:2007xp}, this formula
correctly reproduces the bound-state boundary
S-matrices found in 
\cite{Ahn:2010xa} and \cite{Correa:2009mz}, respectively.  We conclude in Section \ref{sec:discussion} with a
brief discussion of our results.

\section{Bulk fusion}\label{sec:bulk}

We begin by briefly reviewing the new fusion procedure proposed in \cite{Beisert:2015msa}. 
We consider an R-matrix $R(z_1,z_2)$
\be
R(z_1,z_2): \qquad {\cal C}^{n} \otimes {\cal C}^{n} \mapsto   
{\cal C}^{n} \otimes {\cal C}^{n} \,,
\non 
\ee
which is a solution of the (graded) Yang-Baxter equation
\be
R_{12}(z_1,z_2)\, R_{13}(z_1,z_3)\, R_{23}(z_2,z_3) =
R_{23}(z_2,z_3)\, R_{13}(z_1,z_3)\, R_{12}(z_1,z_2) \,,
\label{YBE}
\ee
where $R_{12}(z_1,z_2) = R(z_1,z_2) \otimes \id_{n}\,, R_{13}(z_1,z_3) = {\cal
P}_{23} R_{12}(z_1,z_3) {\cal P}_{23}\,, R_{23}(z_2,z_3) = {\cal
P}_{12} R_{13}(z_2,z_3) {\cal P}_{12}$, and ${\cal P}$ denotes the
(graded) permutation matrix
\be
{\cal P} = \sum_{a,b=1}^{n}  (-1)^{\epsilon_{a} \epsilon_{b}} e_{ab} \otimes e_{ba}\,,
\qquad \left( e_{ab} \right)_{ij} =
\delta_{a,i} \delta_{b,j} \,,
\ee
where $\epsilon_{a} \in \{0, 1\}$ are the gradings.

We further suppose that a bound state forms for 
certain rapidities $(z_1,z_2)$; and correspondingly, $R(z_1,z_2)$ drops 
in rank to $m<n^{2}$, and admits the following 
important decomposition \cite{Beisert:2015msa}
\be
R(z_1,z_2) = \mE(z_{1}, z_{2})\, \mH(z_{1}, z_{2})\, \mF(z_{1}, z_{2})\,, 
\qquad \mF(z_{1}, z_{2})\, \mE(z_{1}, z_{2}) = \id_{m} \,,
\label{decomposition}
\ee
where the matrices act as follows
\be
\mE(z_{1}, z_{2}) &:&  \qquad\ {\cal C}^{m} \mapsto  {\cal C}^{n} \otimes {\cal 
C}^{n} \,, \non \\
\mH(z_{1}, z_{2}) &:&  \qquad\ {\cal C}^{m} \mapsto  {\cal C}^{m} \,, \non \\
\mF(z_{1}, z_{2}) &:& {\cal C}^{n} \otimes {\cal C}^{n} \mapsto   {\cal 
C}^{m}\,. \non 
\ee 
Note that $\mE \mF$ is a projector
\be
\left[\mE(z_{1}, z_{2})\, \mF(z_{1}, z_{2})\right]^{2} = \mE(z_{1}, 
z_{2})\, \mF(z_{1}, z_{2})\,,
\ee
hence (\ref{decomposition}) means that $R(z_1,z_2)$ is ``almost'' 
(i.e., up to the factor $\mH(z_{1},z_{2})$) a 
projector.
It it evident from the decomposition (\ref{decomposition}) that \cite{Beisert:2015msa}
\be
R(z_1,z_2)\, \mE(z_{1}, z_{2})   &=& \mE(z_{1}, z_{2})\, \mH(z_{1}, z_{2})
\,, \label{EH}\\
\mF(z_1,z_2)\, R(z_{1}, z_{2})   &=& \mH(z_{1}, z_{2})\, \mF(z_{1}, z_{2})
\,, \label{HF}\\
R(z_1,z_2)\, \mE(z_{1}, z_{2})\, \mF(z_{1}, z_{2})  &=& R(z_1,z_2) 
\,. \label{REF}
\ee 
These ``fusion identities'' can be used to show that the fused R-matrices \cite{Beisert:2015msa}
\be
R_{\langle 12 \rangle 3}(z_{1}, z_{2}| z_{3}) &=& \mF_{\langle 12 \rangle}(z_{1}, z_{2})\, 
R_{13}(z_{1},z_{3})\, R_{23}(z_{2},z_{3})\, \mE_{\langle 12 
\rangle}(z_{1}, z_{2})  \,, \label{RBA} \\
R_{1 \langle 23 \rangle}(z_{1}| z_{2}, z_{3}) &=& \mF_{\langle 23 
\rangle}(z_{2}, z_{3})\, 
R_{13}(z_{1},z_{3})\, R_{12}(z_{1},z_{2})\, \mE_{\langle 23 
\rangle}(z_{2}, z_{3})  \,, \label{RAB}
\ee
obey corresponding fused (graded) Yang-Baxter equations.
Although these R-matrices are generally not symmetric, they can be made so by a 
similarity transformation \cite{Beisert:2015msa}
\be
R'_{\langle 12 \rangle 3}(z_{1}, z_{2}| z_{3})  &=& W_{\langle 12 \rangle
}(z_{1}, z_{2})\, R_{\langle 12 \rangle 3}(z_{1}, z_{2}| z_{3})\, W^{-1}_{\langle 12 \rangle
}(z_{1}, z_{2}) \,, \label{RpBA} \\
R'_{1 \langle 23 \rangle}(z_{1}| z_{2}, z_{3})  &=& W_{\langle 23 \rangle
}(z_{2}, z_{3})\, R_{1 \langle 23 \rangle}(z_{1}| z_{2}, z_{3})\, W^{-1}_{\langle 23 \rangle
}(z_{2}, z_{3}) \,, \label{RpAB}
\ee
where $W^{T}(z_{1}, z_{2})\, W(z_{1}, z_{2}) = \mH(z_{1}, z_{2})$.

Complementary operators $\bar\mE$ and $\bar\mF$ 
satisfying \cite{Beisert:2015msa}
\be
\mF(z_{1}, z_{2})\, \bar\mE(z_{1}, z_{2}) = 0\,, \quad 
\bar\mF(z_{1}, z_{2})\, \mE(z_{1}, z_{2}) = 0\,, \quad
\bar\mF(z_{1}, z_{2})\, \bar\mE(z_{1}, z_{2}) = \id_{n^{2}-m} \,, 
\label{complopdef}
\ee
as well as the completeness relation
\be
\mE(z_{1}, z_{2})\, \mF(z_{1}, z_{2}) + \bar\mE(z_{1}, z_{2})\, 
\bar\mF(z_{1}, z_{2}) = \id_{n^{2}} \,,
\label{completeness}
\ee
can be used to construct corresponding complementary fused R-matrices
\be
\bar R_{\langle \overline{12} \rangle 3}(z_{1}, z_{2}| z_{3}) &=& \bar\mF_{\langle \overline{12} \rangle}(z_{1}, z_{2})\, 
R_{13}(z_{1},z_{3})\, R_{23}(z_{2},z_{3})\, \bar\mE_{\langle \overline{12} 
\rangle}(z_{1}, z_{2})  \,, \label{compRBA} \\
\bar R_{1 \langle \overline{23} \rangle}(z_{1}| z_{2}, z_{3}) &=& \bar\mF_{\langle \overline{23} 
\rangle}(z_{2}, z_{3})\, 
R_{13}(z_{1},z_{3})\, R_{12}(z_{1},z_{2})\, \bar\mE_{\langle \overline{23} 
\rangle}(z_{2}, z_{3})  \,. \label{compRAB}
\ee

\subsection{AdS/CFT bulk S-matrix: symmetric representation}

Let us now apply this formalism to one copy of the
fundamental $su(2|2)$ AdS/CFT bulk S-matrix. To 
this end, we set
\be
R(z_1,z_2) = S^{AA}(z_1,z_2)
\label{bulkSmat}
\ee
as given by Arutyunov 
and Frolov in \cite{Arutyunov:2008zt}, which  is
reproduced in Appendix \ref{sec:bulkSmatrix}
for the reader's convenience.  This S-matrix satisfies
the graded Yang-Baxter equation (\ref{YBE}) with the gradings 
$\epsilon_{1}=\epsilon_{2}=0\,, \epsilon_{3}=\epsilon_{4}=1$.

We use an elliptic parametrization for the momentum $p$ and the
parameters $x^{\pm}$ for $M$-particle bound states
\cite{Arutyunov:2007tc, Arutyunov:2008zt}
\be
p(z) = 2 \am(z,k) \,, \qquad
x^{\pm}(z)=\frac{M}{2g}\left(\frac{\cn (z,k)}{\sn  (z,k)}\pm
i\right)\left(1+\dn(z,k)\right)\,, \qquad k = - \frac{4g^{2}}{M^{2}} \,,
\ee
such that
\be
\frac{x^{+}}{x^{-}}=e^{i p} \,,
\label{xpxm}
\ee
and
\be
x^{+} + \frac{1}{x^{+}}-x^{-} - \frac{1}{x^{-}} = \frac{2Mi}{g} \,,
\label{shortening}
\ee
where $g>0$ is the coupling constant. However, we henceforth reserve 
$p$ and $x^{\pm}$ for the momentum and parameters of the fundamental particles ($M=1$), and 
$P$ and $y^{\pm}$ for the corresponding quantities of the two-particle bound states ($M=2$).

Consider a pair of fundamental particles with parameters $x^{\pm}_{i} 
= x^{\pm}(z_{i})$, $i=1,2$. These particles form a bound state when 
\cite{Dorey:2006dq, Arutyunov:2007tc} \footnote{We also assume \cite{Arutyunov:2007tc}
that $|x^{\pm}_{i}| > 1$ and $\left(x^{+}_{1}\right)^{*} =  x^{-}_{2}$.}
\be
x^{-}_{1} = x^{+}_{2} \,.
\label{fusioncond}
\ee
Indeed, adding the two constraint equations (\ref{shortening})
\be
x^{+}_{1} + \frac{1}{x^{+}_{1}}-x^{-}_{1} - \frac{1}{x^{-}_{1}} &=& 
\frac{2i}{g} \,, 
\non \\
x^{+}_{2} + \frac{1}{x^{+}_{2}}-x^{-}_{2} - \frac{1}{x^{-}_{2}} &=& 
\frac{2i}{g} \,,
\label{constraints12}
\ee
imposing the fusion condition (\ref{fusioncond}), and making the identifications
\be
y^{+} = x^{+}_{1} \,, \qquad y^{-} = x^{-}_{2} \,,
\label{yparams}
\ee
we arrive at the two-particle bound-state constraint
\be
y^{+} +  \frac{1}{y^{+}}-y^{-} - \frac{1}{y^{-}} = \frac{4i}{g} \,.
\ee
Note that the momentum of the bound state is indeed the sum of the momenta 
of its constituents,  since
\be
e^{i P}=\frac{y^{+}}{y^{-}}=\frac{x^{+}_{1}}{x^{-}_{2}}
=\frac{x^{+}_{1}}{x^{-}_{1}}\frac{x^{+}_{2}}{x^{-}_{2}}=e^{i(p_{1}+p_{2})} \,,
\label{mom}
\ee
where $p_{i} = p(z_{i})$. This bound state lies in the 
8-dimensional symmetric representation of $su(2|2)$ \cite{Bajnok:2008bm}.

When the fusion condition (\ref{fusioncond}) is satisfied, 
the rank of $R(z_1,z_2)$ drops from 16 to 8.
By determining the normalized eigenvectors corresponding to the nonzero eigenvalues, 
we obtain the decomposition (\ref{decomposition}) with
\be
\mE(z_1,z_2) = \tilde \mE(z_1,z_2)\, N^{-1}(z_1,z_2)  \,,
\label{Eresult}
\ee
where
\be
\small
\tilde \mE(z_1,z_2) =
\left(\begin{array}{cccccccc}
	   0 & 0 & 1 & 0 & 0 & 0 & 0 & 0 \\
	   0 & \frac{1}{\sqrt{2}} & 0 & -\frac{1}{\sqrt{2}}a_{2} & 0 & 0 & 0 & 0 \\
	   0 & 0 & 0 & 0 & 0 & 0 & 0 & a_{5} \\
	   0 & 0 & 0 & 0 & 0 & a_{5} & 0 & 0 \\
	   0 & \frac{1}{\sqrt{2}} & 0 & \frac{1}{\sqrt{2}}a_{2} & 0 & 0 & 0 & 0 \\
	   1 & 0 & 0 & 0 & 0 & 0 & 0 & 0 \\
	   0 & 0 & 0 & 0 & 0 & 0 & a_{5} & 0 \\
	   0 & 0 & 0 & 0 & a_{5} & 0 & 0 & 0 \\
	   0 & 0 & 0 & 0 & 0 & 0 & 0 & a_{10} \\
	   0 & 0 & 0 & 0 & 0 & 0 & a_{10} & 0 \\
	   0 & 0 & 0 & 0 & 0 & 0 & 0 & 0 \\
	   0 & 0 & 0 & -\sqrt{2} a_{8} & 0 & 0 & 0 & 0 \\
	   0 & 0 & 0 & 0 & 0 & a_{10} & 0 & 0 \\
	   0 & 0 & 0 & 0 & a_{10} & 0 & 0 & 0 \\
	   0 & 0 & 0 & \sqrt{2} a_{8} & 0 & 0 & 0 & 0 \\
	   0 & 0 & 0 & 0 & 0 & 0 & 0 & 0 \\
       \end{array}\right) \,,
\ee 
and $N(z_1,z_2)$ is the diagonal matrix 
\be
N(z_1,z_2) = \diag(1\,,1\,,1\,,n_{1}\,,n_{2}\,,n_{2}\,,n_{2}\,,n_{2}) \,, 
\quad
n_{1}=\sqrt{a_{2}^{2}+4a_{8}^{2}}\,, \quad 
n_{2}=\sqrt{a_{5}^{2}+a_{10}^{2}}\,,
\label{Nmat}
\ee
where the $a_{k}=a_{k}(z_{1},z_{2})$ are given by (\ref{ak}). 
Moreover,
\be
\mF(z_1,z_2) = \mE^{T}(z_1,z_2) = N^{-1}(z_1,z_2) \, \tilde \mE^{T}(z_1,z_2) = 
N^{-1}(z_1,z_2) \, \tilde \mF(z_1,z_2) \,,
\label{Fresult}
\ee
where we have defined $\tilde \mF(z_1,z_2) = \tilde \mE^{T}(z_1,z_2)$. 
Finally,
\be
\mH(z_1,z_2) = \diag(1\,,1\,,1\,,h_{1}\,,h_{2}\,,h_{2}\,,h_{2}\,,h_{2}) \,, 
\quad
h_{1} = a_{2}+a_{4}\,, \quad 
h_{2} = a_{5}+a_{6}\,.
\ee

Performing the similarity transformation (\ref{RpAB}) with the matrix
\be
\small
W(z_1,z_2) =
\left(\begin{array}{cccccccc}
	   0 & 0 & 1 & 0 & 0 & 0 & 0 & 0 \\
	   0 & 1 & 0 & 0 & 0 & 0 & 0 & 0 \\
	   1 & 0 & 0 & 0 & 0 & 0 & 0 & 0 \\
	   0 & 0 & 0 & w_{1} & 0 & 0 & 0 & 0 \\
	   0 & 0 & 0 & 0 & 0 & 0 & 0 & w_{2} \\
	   0 & 0 & 0 & 0 & 0 & w_{2} & 0 & 0 \\
	   0 & 0 & 0 & 0 & 0 & 0 & w_{2} & 0 \\
	   0 & 0 & 0 & 0 & w_{2} & 0 & 0 & 0 \\
       \end{array}\right)\,, \quad w_{1}=\sqrt{a_{2}+a_{4}}\,, \quad 
       w_{2}=\sqrt{a_{5}+a_{6}} \,,
\label{Wmat}
\ee  
we obtain
\be
R'_{1 \langle 23 \rangle}(z_{1}| z_{2}, z_{3})  = U_{\langle 23 \rangle}(z_{2}, z_{3})\, 
\tilde \mF_{\langle 23 
\rangle}(z_{2}, z_{3})\, 
R_{13}(z_{1},z_{3})\, R_{12}(z_{2},z_{2})\, \tilde \mE_{\langle 23 
\rangle}(z_{2}, z_{3})\, V^{-1}_{\langle 23 \rangle}(z_{2}, z_{3}) \,, 
\label{sim}
\ee
where we have defined the new matrices $U$ and $V$, which evidently 
have the same matrix structure as $W$, but have different 
matrix elements
\be
U(z_1,z_2) &=& W(z_1,z_2)\, N^{-1}(z_1,z_2) = 
W(z_1,z_2)\Big\vert_{w_{i} \rightarrow u_{i}}\,,  
\non \\
V(z_1,z_2) &=& W(z_1,z_2)\, N(z_1,z_2)= 
W(z_1,z_2)\Big\vert_{w_{i} \rightarrow v_{i}} \,.
\label{UVdefs}
\ee
By explicit computation we obtain the following results for these matrix elements  
\be
u_{1}&=& \frac{w_{1}}{n_{1}} = \left[2 i g \sin(p_{1}/2) 
\sin(p_{2}/2) \right]^{-1}\,, \non\\
u_{2}&=& \frac{w_{2}}{n_{2}} = e^{-i p_{1}/2} 
\frac{\eta(z_{12},2)}{\eta(z_{2},1)}\,, \non\\
v_{1}&=& w_{1} n_{1} = 8 i g 
\frac{\sin^{2}(p_{1}/2)\sin^{2}(p_{2}/2)}{\sin^{2}((p_{1}+p_{2})/2)}(1-g^{2}\sin(p_{1}/2) 
\sin(p_{2}/2)\sin^{2}((p_{1}+p_{2})/2)\,, \non\\
v_{2}&=& w_{2} n_{2} = 
\left[\left(1+e^{-i(p_{1}+p_{2})/2}\right)^{-1}+ e^{i 
p_{1}}\left(1+e^{i(p_{1}+p_{2})/2}\right)^{-1} \right] \frac{\eta(z_{2},1)}{\eta(z_{12},2)}\,,  \label{defuv}
\ee 
where $\eta(z, M)$ is defined in (\ref{eta}), and the rapidity 
$z_{12}$ is defined such that 
\be
y^{+}(z_{12}) = x^{+}(z_{1})\,, \qquad
y^{-}(z_{12}) = x^{-}(z_{2})\,,
\ee
as in (\ref{yparams}). Remarkably, the square roots in $u_{i}$ and 
$v_{i}$ (recall the definitions of $n_{i}$ and $w_{i}$ given in 
(\ref{Nmat}) and (\ref{Wmat})) have all disappeared.

Using these results to evaluate (\ref{sim}), we have verified 
numerically that this 
fused R-matrix coincides with $S^{AB}$ in \cite{Arutyunov:2008zt} 
\footnote{As noted in \cite{Ahn:2010xa}, there are two
typos in the coefficients of $S^{AB}$ listed in Section 6.1.2 of
\cite{Arutyunov:2008zt}.  In $a_{13}$, the factor in the numerator $(x_1^- - y_2^+)$
should be instead $(x_1^+ - y_2^+)$; i.e., the $x_1^-$ should be
changed to $x_1^+$.  And in $a_{14}$, the factor in the numerator
$(1-y_2^- x_1^-)$ should be instead $(1-y_2^- x_1^+)$ ; i.e., the
$x_1^-$ should be changed to $x_1^+$.}
\be
S^{AB}(z_{1},z_{23}) =  R'_{1 \langle 23 \rangle}(z_{1}| z_{2}, z_{3})  \,.
\label{SAB}
\ee
A similar result was argued in \cite{Beisert:2015msa}.

\subsection{AdS/CFT bulk S-matrix: antisymmetric representation}

We now proceed to construct the complementary fused S-matrix 
(\ref{compRAB}), which corresponds to the antisymmetric 
representation of $su(2|2)$ \cite{Bajnok:2008bm}, which is also 
8-dimensional.
The required complementary operators $\bar\mE$ and $\bar\mF$ can be obtained 
by considering the ``opposite'' fusion condition 
\be
x^{+}_{1} = x^{-}_{2} \,.
\label{antisymfusioncond}
\ee
Since all the $a_{k}$ (\ref{ak}) except $a_{1}$ have a simple pole at 
this point, it is convenient to introduce rescaled quantities 
$\hat a_{k}(z_{1},z_{2}) = (x^{+}_{1} - x^{-}_{2}) 
a_{k}(z_{1},z_{2})$ and $\hat R(z_1,z_2) = (x^{+}_{1} - 
x^{-}_{2})\, S^{AA}(z_1,z_2)$.

When the fusion condition (\ref{antisymfusioncond}) is satisfied, 
the rank of $\hat R(z_1,z_2)$ indeed drops from 16 to 8, and we obtain
the decomposition 
\be
\hat R(z_1,z_2) = \mE_{\cal A}(z_{1}, z_{2})\, \mH_{\cal A}(z_{1}, z_{2})\, 
\mF_{\cal A}(z_{1}, z_{2})\,, 
\qquad \mF_{\cal A}(z_{1}, z_{2})\, \mE_{\cal A}(z_{1}, z_{2}) = 
\id_{8} \,,
\label{antisymdecomposition}
\ee
with
\be
\mE_{\cal A}(z_1,z_2) = \tilde \mE_{\cal A}(z_1,z_2)\, N^{-1}(z_1,z_2)  \,,
\label{antisymEresult}
\ee
where
\be
 \tilde \mE_{\cal A}(z_1,z_2) =
\left(\begin{array}{cccccccc}
0 & 0 & 0 & 0 & 0 & 0 & 0 & 0 \\
0 & 0 & 0 & -\frac{1}{\sqrt{2}}a_{2} & 0 & 0 & 0 & 0 \\
0 & 0 & 0 & 0 & 0 & 0 & 0 & a_{5} \\
0 & 0 & 0 & 0 & 0 & a_{5} & 0 & 0 \\
0 & 0 & 0 & \frac{1}{\sqrt{2}}a_{2} & 0 & 0 & 0 & 0 \\
0 & 0 & 0 & 0 & 0 & 0 & 0 & 0 \\
0 & 0 & 0 & 0 & 0 & 0 & a_{5} & 0 \\
0 & 0 & 0 & 0 & a_{5} & 0 & 0 & 0 \\
0 & 0 & 0 & 0 & 0 & 0 & 0 & a_{10} \\
0 & 0 & 0 & 0 & 0 & 0 & a_{10} & 0 \\
0 & 0 & 1 & 0 & 0 & 0 & 0 & 0 \\
0 & \frac{1}{\sqrt{2}} & 0 & -\sqrt{2}a_{8} & 0 & 0 & 0 & 0 \\
0 & 0 & 0 & 0 & 0 & a_{10} & 0 & 0 \\
0 & 0 & 0 & 0 & a_{10} & 0 & 0 & 0 \\
0 & \frac{1}{\sqrt{2}} & 0 & \sqrt{2} a_{8} & 0 & 0 & 0 & 0 \\
1 & 0 & 0 & 0 & 0 & 0 & 0 & 0 \\
\end{array}\right) \,,
\label{EtildeA}
\ee 
and $N(z_1,z_2)$ is again given by (\ref{Nmat}).
(Note that the singular factors
in $N(z_1,z_2)$ and $\tilde \mE_{\cal A}(z_1,z_2)$ are canceled in $\mE_{\cal A}(z_1,z_2)$.) Moreover, 
$\mF_{\cal A}(z_1,z_2) = \mE^{T}_{\cal A}(z_1,z_2)$, and
\be
\mH_{\cal A}(z_1,z_2) = \diag(\hat a_{3}\,,\hat a_{3}\,, \hat a_{3}\,,\hat h_{1}\,,\hat h_{2}\,,\hat h_{2}\,,\hat h_{2}\,,\hat h_{2}) \,, 
\quad
\hat h_{1} = \hat a_{2}+\hat a_{4}\,, \quad 
\hat h_{2} = \hat a_{5}+\hat a_{6}\,.
\ee
Finally, the complementary operators are given by \cite{Beisert:2015msa}
\be
\bar \mE(z_{1}\,, z_{2}) = {\cal P} \mE_{\cal A}(z_{2}\,, z_{1}) \,, \qquad  
\bar \mF(z_{1}\,, z_{2}) = \bar \mE^{T}(z_{1}\,, z_{2}) \,,
\label{complemops}
\ee  
where it is now understood that $z_{1}$ and $z_{2}$ correspond to the 
original fusion condition (\ref{fusioncond}). These complementary operators, 
together with the original operators (\ref{Eresult}) and 
(\ref{Fresult}),
satisfy the relations (\ref{complopdef}) and  (\ref{completeness}).

We have verified numerically that the complementary fused R-matrix 
obtained following (\ref{compRAB}), up to a similarity transformation, is proportional 
to the complex conjugate of $S^{AB}$ in \cite{Arutyunov:2008zt} 
\be
\left[\frac{S^{AB}(z_{1},z_{23})}{a_{3}(z_{1},z_{2}) 
a_{3}(z_{1},z_{3})}\right]^{*} =  \bar R'_{1 \langle \overline{23} 
\rangle}(z_{1}| z_{2}, z_{3})  \,,
\label{SABconj}
\ee
as expected for the antisymmetric representation \cite{Bajnok:2008bm}. Again, a similar 
result was obtained in \cite{Beisert:2015msa}.

\section{Boundary fusion}\label{sec:boundary}

We now generalize the above discussion to the case of boundary 
scattering. Let $K(z)$ 
\be
K(z): \qquad {\cal C}^{n} \mapsto {\cal C}^{n} 
\non 
\ee
be a solution of the boundary Yang-Baxter equation
\cite{Cherednik:1985vs, Sklyanin:1988yz, Ghoshal:1993tm}
\be
R_{12}(z_{1},z_{2})\, K_{1}(z_{1})\, R_{21}(z_{2},-z_{1})\,
K_{2}(z_{2}) = K_{2}(z_{2})\, R_{12}(z_{1},-z_{2})\,
K_{1}(z_{1})\, R_{21}(-z_{2},-z_{1})\,,
\label{BYBE}
\ee
where $R_{21}(z_{1},z_{2}) = {\cal P}_{12}\, R_{12}(z_{1},z_{2})\,
{\cal P}_{12}$, $K_{1}(z) = K(z) \otimes \id_{n}$ and
$K_{2}(z) = {\cal P}_{12}\, K_{1}(z)\,
{\cal P}_{12}$.

We propose that the fused K-matrix is given by (cf. Eq. 
(3.5) in \cite{Mezincescu:1991ke})
\be
K_{\langle 12 \rangle}(z_{1}, z_{2}) &=& \mF_{\langle 12 \rangle}(z_{1}, z_{2})\, 
K_{1}(z_{1})\, R_{21}(z_{2},-z_{1})\, K_{2}(z_{2})\, {\cal P}_{12}
\mE_{\langle 12 \rangle}(-z_{2},- z_{1})  \,. \label{Kfused} 
\ee
Indeed, we show in Appendix \ref{sec:proof} that this object satisfies 
the fused boundary Yang-Baxter equation
\be
\lefteqn{R_{\langle 12 \rangle 3}(z_{1},z_{2}|z_{3})\, K_{\langle 12 
\rangle}(z_{1}, z_{2})\, R_{3 \langle 12 \rangle}(z_{3}| 
-z_{2}, -z_{1})\, K_{3}(z_{3})} \non \\
&&= K_{3}(z_{3})\, R_{\langle 12 \rangle 3}(z_{1},z_{2}| -z_{3})\,
K_{\langle 12 \rangle}(z_{1}, z_{2})\,
R_{3 \langle 12 \rangle}(-z_{3}| -z_{2}, -z_{1}) \,,
\label{fusedBYBE}
\ee
where $R_{\langle 12 \rangle 3}$ and $R_{1 \langle 23 \rangle}$ are 
given by (\ref{RBA}) and (\ref{RAB}), respectively. The boundary 
fusion formula (\ref{Kfused}) is the main result of this paper.
Performing a similarity transformation $\mE \rightarrow \mE\, W^{-1}\,, \mF 
\rightarrow  W\, \mF$ as in (\ref{RpBA}) and (\ref{RpAB}) gives
\be
K'_{\langle 12 \rangle}(z_{1}, z_{2}) = W_{\langle 12 \rangle}(z_{1}, z_{2})\, 
K_{\langle 12 \rangle}(z_{1}, z_{2})\, W^{-1}_{\langle 12 \rangle}(z_{1}, z_{2}) 
\,.
\label{Kpfused}
\ee

Using the complementary operators $\bar\mE$ and $\bar\mF$ 
satisfying (\ref{complopdef}) and (\ref{completeness}), complementary 
fused boundary K-matrices can be constructed in a similar manner
\be
\bar K_{\langle \overline{12} \rangle}(z_{1}, z_{2}) &=& \bar \mF_{\langle \overline{12} \rangle}(z_{1}, z_{2})\, 
K_{1}(z_{1})\, R_{21}(z_{2},-z_{1})\, K_{2}(z_{2})\, {\cal P}_{12}
\bar \mE_{\langle \overline{12} \rangle}(-z_{2},- z_{1})  \,. \label{compKfused} 
\ee
The proof of this result is sketched in Appendix 
\ref{sec:proof}.

\subsection{AdS/CFT boundary S-matrix: symmetric representation}

Let us illustrate the boundary fusion formula (\ref{Kpfused}) with the simplest AdS/CFT boundary S-matrix 
\be
K(z) = \diag(e^{-i p/2}\,, -e^{i p/2} \,, 1\,, 1 ) \,,
\label{RBSM}
\ee
corresponding to a $Y=0$ brane \cite{Hofman:2007xp}. Using our 
previous expressions for $\mE$ and $\mF$ (\ref{Eresult}), (\ref{Fresult}), we 
obtain (cf. (\ref{sim}))
\be
K'_{\langle 12 \rangle}(z_{1}, z_{2}) =  U_{\langle 12 \rangle}(z_{1}, z_{2})\, 
\tilde \mF_{\langle 12 \rangle}(z_{1}, z_{2})\, 
K_{1}(z_{1})\, R_{21}(z_{2},-z_{1})\, K_{2}(z_{2})\, {\cal P}_{12}
\tilde \mE_{\langle 12 \rangle}(-z_{2},- z_{1})
T^{-1}_{\langle 12 \rangle}(z_{1},z_{2})\,, \non \\
\label{Ksim}
\ee
where $U$ is defined in (\ref{UVdefs}), and $T$ is the following new matrix
\be
T(z_1,z_2) &=& W(z_1,z_2)\, N(-z_2, -z_1) = 
W(z_1,z_2)\Big\vert_{w_{i} \rightarrow t_{i}} \,,
\ee
which also has the same matrix structure as $W$, but has different 
matrix elements. We find that these matrix elements are given by
\be
t_{1} &=& w_{1}(z_1,z_2)\, n_{1}(-z_2, -z_1) = v_{1}\,, \non \\
t_{2} &=& w_{2}(z_1,z_2)\, n_{2}(-z_2, -z_1) = e^{-i p_{1}/2}
\frac{\eta(z_{1},1)}{\eta(z_{2},1)}v_2\,,
\ee 
where $v_1$ and $v_2$ are given in (\ref{defuv}).
Using these results to evaluate (\ref{Ksim}), we have verified that this 
fused K-matrix coincides (up to an overall scalar factor) with the 
bound-state $Y=0$ boundary S-matrix $R^{B}$ in \cite{Ahn:2010xa},
\be
R^{B}(z_{12}) & = & e^{i P/2}\, K'_{\langle 12 \rangle}(z_{1}, 
z_{2})  \non \\
& = & \left( \begin{array}{cccccccc}
r_{1} & 0 & 0 & 0 & 0 & 0 & 0 & 0 \\
0 & r_{2} & 0 & r_{5} & 0 & 0 & 0 & 0 \\
0 & 0 & r_{3} & 0 & 0 & 0 & 0 & 0\\
0 & r_{6} & 0 & r_{4} & 0 & 0 & 0 & 0\\
0 & 0 & 0 & 0 & r_{7} & 0 & 0 & 0\\
0 & 0 & 0 & 0 & 0 & r_{7} & 0 & 0\\
0 & 0 & 0 & 0 & 0 & 0 & r_{8} & 0\\
0 & 0 & 0 & 0 & 0 & 0 & 0 & r_{8}
\end{array} \right) \,,
\label{RB}
\ee
where
\be
r_{1} &=& 1 \,, \qquad 
r_{2} = -\frac{\frac{1}{y^{-}}+y^{-}}{\frac{1}{y^{+}}+y^{-}} \,, \qquad
r_{3} = e^{i P} \,,  \qquad
r_{4} = \frac{\frac{1}{y^{+}}+y^{+}}{\frac{1}{y^{+}}+y^{-}}\,, \non \\
r_{5} &=& - r_{6} =  e^{i P/2} \frac{y^{-}-y^{+}}{1 + y^{-} y^{+}} \,, \qquad 
\quad \
r_{7} = - r_{8} = e^{i P/2} \,.
\ee
While the verification of some of the matrix elements is straightforward 
(e.g., $r_{3}$ requires just (\ref{mom}), and $r_{2}$ requires use of 
(\ref{fusioncond}) and (\ref{constraints12})), those involving 
$\eta$'s are much more complicated. Nevertheless, by using the 
expression for $\eta$ in terms of a square root (\ref{eta}) and using ${\tt PowerExpand}$ in 
${\tt Mathematica}$, we managed to explicitly check all of the matrix 
elements. 

\subsection{AdS/CFT boundary S-matrix: antisymmetric representation}

Computing the complementary fused K-matrix (\ref{compKfused}) using 
the complementary operators (\ref{complemops}), as well as the fundamental 
bulk (\ref{bulkSmat}) and boundary (\ref{RBSM}) S-matrices, we obtain
the diagonal matrix
\be
\bar K_{\langle \overline{12} \rangle}(z_{1}, z_{2}) = 
-\frac{\cos(p_{2}/2)}{\cos(p_{1}/2)}\diag(1\,, 1\,, 1\,, 
-1\,, -e^{i P/2}\,, e^{-i P/2}\,, -e^{i P/2}\,,  e^{-i P/2}) \,,
\label{CY}
\ee
which satisfies the fused boundary Yang-Baxter equation
(\ref{fusedBYBE}) with the complementary fused R-matrices $\bar 
R_{\langle \overline{12} \rangle 3}$ and $\bar R_{3 \langle \overline{12} \rangle}$.
The result (\ref{CY}) can be related by a similarity transformation, 
up to an overall scalar factor, to the antisymmetric 
representation $M=2$ bound-state boundary S-matrix $\mathcal{R}_2$ obtained in 
\cite{Correa:2009mz}.

\section{Discussion}\label{sec:discussion}

We have found a fusion formula (\ref{Kfused}) that is applicable to
AdS/CFT boundary S-matrices, many examples of which are now known.  We
have focused on the $Y=0$ example only for simplicity.  Although we have
used the fusion formula to obtain only the $M=2$ bound-state $Y=0$ 
boundary S-matrices, we expect that a further generalization (along the 
lines of \cite{Zhou:1995zy}) is possible for recovering the higher ($M>2$)
bound-state boundary S-matrices found in \cite{Correa:2009mz} and 
\cite{Palla:2011eu} for antisymmetric and symmetric representations, 
respectively.

We have noticed that the expressions generated by both the bulk and
boundary fusion formulas are generally very complicated, and require
considerable effort to simplify, particularly in the symmetric
representation.  It would be interesting to find a more efficient way
of writing the basic elements (R, $\mE$ and $\mF$) that lead directly
to simpler results for the fused quantities.

\section*{Acknowledgments}
The work of RN was supported in part by the National Science
Foundation under Grant PHY-1212337, and by a Cooper fellowship.
RP thanks the Sao Paulo Research Foundation (FAPESP),
grants \# 2014/00453-8 and \# 2014/20364-0, for financial support.

\appendix

\section{Fundamental bulk S-matrix}\label{sec:bulkSmatrix} 

The graded bulk S-matrix for a pair of particles in 
the fundamental (4-dimensional) representation is 
given by \cite{Arutyunov:2008zt}
\be
S^{AA}(z_1,z_2) = \sum_{k=1}^{10}a_{k}(z_1,z_2) \Lambda_{k}\,,
\label{bulkS}
\ee
where the $16 \times 16$ matrices $\Lambda_{1}\,, \ldots\,, \Lambda_{10}$
are given in terms of quantities $E_{kilj}$ defined by
\be
E_{kilj} = e_{ki} \otimes e_{lj} \,, 
\label{Ekilj}
\ee
with indices that run from 1 to 4.

Hence, $S^{AA}(z_1,z_2)$ has the following matrix structure
\be
\left(
\begin{array}{cccccccccccccccc}
 a_1 & 0 & 0 & 0 & 0 & 0 & 0 & 0 & 0 & 0 & 0 & 0 & 0 & 0 & 0 & 0 \\
 0 & \frac{a_1}{2}+\frac{a_2}{2} & 0 & 0 & \frac{a_1}{2}-\frac{a_2}{2} & 0 & 0 & 0 & 0 & 0 & 0 & a_7 & 0 & 0 &
   -a_7 & 0 \\
 0 & 0 & a_5 & 0 & 0 & 0 & 0 & 0 & a_9 & 0 & 0 & 0 & 0 & 0 & 0 & 0 \\
 0 & 0 & 0 & a_5 & 0 & 0 & 0 & 0 & 0 & 0 & 0 & 0 & a_9 & 0 & 0 & 0 \\
 0 & \frac{a_1}{2}-\frac{a_2}{2} & 0 & 0 & \frac{a_1}{2}+\frac{a_2}{2} & 0 & 0 & 0 & 0 & 0 & 0 & -a_7 & 0 & 0 &
   a_7 & 0 \\
 0 & 0 & 0 & 0 & 0 & a_1 & 0 & 0 & 0 & 0 & 0 & 0 & 0 & 0 & 0 & 0 \\
 0 & 0 & 0 & 0 & 0 & 0 & a_5 & 0 & 0 & a_9 & 0 & 0 & 0 & 0 & 0 & 0 \\
 0 & 0 & 0 & 0 & 0 & 0 & 0 & a_5 & 0 & 0 & 0 & 0 & 0 & a_9 & 0 & 0 \\
 0 & 0 & a_{10} & 0 & 0 & 0 & 0 & 0 & a_6 & 0 & 0 & 0 & 0 & 0 & 0 & 0 \\
 0 & 0 & 0 & 0 & 0 & 0 & a_{10} & 0 & 0 & a_6 & 0 & 0 & 0 & 0 & 0 & 0 \\
 0 & 0 & 0 & 0 & 0 & 0 & 0 & 0 & 0 & 0 & a_3 & 0 & 0 & 0 & 0 & 0 \\
 0 & a_8 & 0 & 0 & -a_8 & 0 & 0 & 0 & 0 & 0 & 0 & \frac{a_3}{2}+\frac{a_4}{2} & 0 & 0 &
   \frac{a_3}{2}-\frac{a_4}{2} & 0 \\
 0 & 0 & 0 & a_{10} & 0 & 0 & 0 & 0 & 0 & 0 & 0 & 0 & a_6 & 0 & 0 & 0 \\
 0 & 0 & 0 & 0 & 0 & 0 & 0 & a_{10} & 0 & 0 & 0 & 0 & 0 & a_6 & 0 & 0 \\
 0 & -a_8 & 0 & 0 & a_8 & 0 & 0 & 0 & 0 & 0 & 0 & \frac{a_3}{2}-\frac{a_4}{2} & 0 & 0 &
   \frac{a_3}{2}+\frac{a_4}{2} & 0 \\
 0 & 0 & 0 & 0 & 0 & 0 & 0 & 0 & 0 & 0 & 0 & 0 & 0 & 0 & 0 & a_3 \\
\end{array}
\right) \non 
\ee
and the matrix elements $a_{k} = a_{k}(z_1,z_2)$ are given by \cite{Arutyunov:2008zt}
{\allowdisplaybreaks 
\be
a_1 &=& 1\,, \non \\
a_2 &=& 2\,\frac{(x^+_1-x^+_2) (x^-_1
   x^+_2-1)x^-_2}{(x^+_1 - x^-_2)(x^-_1 x^-_2-1)
    x^+_2}-1\,, \non \\
a_3 &=& \frac{x^+_2-x^-_1}{x^-_2-x^+_1}
\frac{\tilde{\eta}_1
  \tilde{\eta}_2}{ \eta_1\eta_2} \,, \non \\
a_4 &=& \frac{(x^-_1-x^+_2)}{(x^-_2-x^+_1)}\frac{\tilde{\eta}_1
 \tilde{\eta}_2}{ \eta_1\eta_2} - 2\,\frac{
   (x^-_2 x^+_1-1) (x^+_1-x^+_2)
  x^-_1 }{(x^-_1 x^-_2-1)
   (x^-_2-x^+_1) x^+_1} \frac{\tilde{\eta}_1
  \tilde{\eta}_2}{ \eta_1\eta_2} \,, \non \\
a_5 &=& \frac{x^-_2-x^-_1}{x^-_2-x^+_1 } \frac{\tilde{\eta}_2}{\eta 
_2}  \,, \non \\
a_6 &=& 
\frac{x^+_1-x^+_2}{x^+_1-x^-_2}\frac{\tilde{\eta}_1}{\eta_1}\,, \non 
\\
a_7 &=& -\frac{i (x^-_1-x^+_1)
   (x^-_2-x^+_2)
   (x^+_1-x^+_2)}{(x^-_1
   x^-_2-1) (x^-_2-x^+_1) }\frac{1}{\eta
   _1 \eta _2}\,, \non \\
a_8 &=& \frac{i x^-_1 x^-_2
   (x^+_1-x^+_2)}{(x^-_1 x^-_2-1)
   (x^-_2-x^+_1) x^+_1
   x^+_2}\tilde{\eta}_1\tilde{\eta}_2 \,, \non \\
a_9 &=& \frac{x^+_1-x^-_1}{x^+_1-x^-_2}\frac{\tilde{\eta}_2}{\eta_1} 
\,, \non \\
a_{10} &=& 
\frac{x^-_2-x^+_2}{x^-_2-x^+_1}\frac{\tilde{\eta}_1}{\eta_2}\,.
\label{ak}
\ee
}
Moreover, 
\be
\eta_1 = e^{ip_2/2}\eta(z_1)\,,\quad 
\eta_2 =\eta(z_2)\,,\quad
\tilde{\eta}_1 = \eta(z_1)\,,\quad 
\tilde{\eta}_2 = e^{ip_1/2}\eta(z_2)\,,
\ee
where $\eta(z) = \eta(z, 1)$, with
\be
\eta(z, M) = e^{i p/4} \sqrt{i(x^{-}-x^{+})} 
= \sqrt{\frac{2M}{g}}\frac{\dn\, \frac{z}{2}\big(
\cn\, \frac{z}{2}+i \, \sn\, \frac{z}{2}\dn\,
\frac{z}{2}\big)}{1+\frac{4g^2}{M^{2}}\, {\sn^4\frac{z}{2}}}\,.
\label{eta}
\ee

\section{Proof of the boundary fusion formulas}\label{sec:proof} 

We first show here that the fused K-matrix (\ref{Kfused}) satisfies the fused
boundary Yang-Baxter equation (\ref{fusedBYBE}).
{\allowdisplaybreaks 
We use here the following shorthand notation,
\be
&&R_{12}(z_1,z_2)=R_{12},\quad R_{13}(z_1,z_3)=R_{13},\quad R_{23}(z_2,z_3)=R_{23},\nonumber\\
&&\mE_{\ud}(z_1,z_2)=\mE_{\ud},\quad\mF_{\ud}(z_1,z_2)=\mF_{\ud},\quad \mH_{\ud}(z_1,z_2)=\mH_{\ud} \,, \nonumber 
\ee
and
\be
&&R_{21}(z_2,-z_1)=R_{21}, \quad R_{31}(z_3,-z_1)=R_{31}, \quad R_{32}(z_3,-z_2)=R_{32},\nonumber\\
&&K_1(z_1)=K_1,\quad K_2(z_2)=K_2,\quad K_3(z_3)=K_3 \,. \nonumber 
\ee
If the arguments of a given matrix do not fit the above notation we write them explicitly. 
\be
&&\mH_{\ud}(z_1,z_2)R_{\ud 3}(z_1,z_2|z_3)K_{\ud}(z_1,z_2)R_{3\langle 12\rangle}(z_3|-z_2,-z_1)K_{3}(z_3)
\nonumber\\
&&=\underbrace{\mH_{\ud}\mF_{\ud}}_{\textrm{Eq.}(\ref{HF})}R_{13}R_{23}\mE_{\ud}\mF_{\ud}K_1R_{21}K_2\mP_{12}\mE_{\ud}(-z_2,-z_1)\mF_{\ud}(-z_2,-z_1)R_{32}(z_3,-z_1)\cdots\nonumber\\
&&\qquad\qquad \cdots R_{31}(z_3,-z_2)\underbrace{\mE_{\ud}(-z_2,-z_1)K_3}_{\textrm{commute}}
\nonumber\\
&&=\mF_{\ud}\underbrace{R_{12}R_{13}R_{23}}_{\textrm{Eq.}(\ref{YBE})}\mE_{\ud}\mF_{\ud}K_1R_{21}K_2\mP_{12}\mE_{\ud}(-z_2,-z_1)\mF_{\ud}(-z_2,-z_1)R_{32}(z_3,-z_1)\cdots\nonumber\\
&&\qquad\qquad \cdots R_{31}(z_3,-z_2)K_3\mE_{\ud}(-z_2,-z_1)
\nonumber\\
&&=\mF_{\ud}R_{23}R_{13}\underbrace{R_{12}\mE_{\ud}\mF_{\ud}}_{\textrm{Eq.}(\ref{REF})}K_1R_{21}K_2\mP_{12}\mE_{\ud}(-z_2,-z_1)\mF_{\ud}(-z_2,-z_1)R_{32}(z_3,-z_1)\cdots\nonumber\\
&&\qquad\qquad \cdots R_{31}(z_3,-z_2)K_3\mE_{\ud}(-z_2,-z_1)
\nonumber\\
&&=\mF_{\ud}R_{23}R_{13}\underbrace{R_{12}K_1R_{21}K_2}_{\textrm{Eq.}(\ref{BYBE})}\mP_{12}\mE_{\ud}(-z_2,-z_1)\mF_{\ud}(-z_2,-z_1)R_{32}(z_3,-z_1)\cdots\nonumber\\
&&\qquad\qquad \cdots R_{31}(z_3,-z_2)K_3\mE_{\ud}(-z_2,-z_1)
\nonumber\\
&&=\mF_{\ud}R_{23}\underbrace{R_{13}K_2}_{\textrm{commute}}R_{12}(z_1,-z_2)K_1\underbrace{R_{21}(-z_2,-z_1)\mP_{12}}_{\mP_{12}^2=1}\mE_{\ud}(-z_2,-z_1)\mF_{\ud}(-z_2,-z_1)\cdots\nonumber\\
&&\qquad\qquad \cdots R_{32}(z_3,-z_1)R_{31}(z_3,-z_2)K_3\mE_{\ud}(-z_2,-z_1)
\nonumber\\
&&=\mF_{\ud}R_{23}K_2R_{13}R_{12}(z_1,-z_2)K_1\mP_{12}\underbrace{R_{12}(-z_2,-z_1)\mE_{\ud}(-z_2,-z_1)\mF_{\ud}(-z_2,-z_1)}_{\textrm{Eq.}(\ref{REF})}\cdots\nonumber\\
&&\qquad\qquad \cdots R_{32}(z_3,-z_1)R_{31}(z_3,-z_2)K_3\mE_{\ud}(-z_2,-z_1)
\nonumber\\
&&=\mF_{\ud}R_{23}K_2R_{13}R_{12}(z_1,-z_2)K_1\mP_{12}\underbrace{R_{12}(-z_2,-z_1)R_{32}(z_3,-z_1)R_{31}(z_3,-z_2)}_{\textrm{Eq.}(\ref{YBE})}K_3\cdots\nonumber\\
&&\qquad\qquad \cdots \mE_{\ud}(-z_2,-z_1)
\nonumber\\
&&=\mF_{\ud}R_{23}K_2R_{13}R_{12}(z_1,-z_2)K_1\underbrace{\mP_{12}R_{31}(z_3,-z_2)R_{32}(z_3,-z_1)R_{12}(-z_2,-z_1)}_{\mP_{12}^2=1}K_3\cdots\nonumber\\
&&\qquad\qquad \cdots \mE_{\ud}(-z_2,-z_1)
\nonumber\\
&&=\mF_{\ud}R_{23}K_2R_{13}R_{12}(z_1,-z_2)\underbrace{K_1R_{32}}_{\textrm{commute}}R_{31}\underbrace{R_{21}(-z_2,-z_1)\mP_{12}K_3}_{\textrm{commute}} \mE_{\ud}(-z_2,-z_1)
\nonumber\\
&&=\mF_{\ud}R_{23}K_2\underbrace{R_{13}R_{12}(z_1,-z_2)R_{32}}_{\textrm{Eq.}(\ref{YBE})}K_1R_{31}K_3R_{21}(-z_2,-z_1)\mP_{12}\mE_{\ud}(-z_2,-z_1)
\nonumber\\
&&=\mF_{\ud}R_{23}K_2R_{32}R_{12}(z_1,-z_2)\underbrace{R_{13}K_1R_{31}K_3}_{\textrm{Eq.}(\ref{BYBE})}R_{21}(-z_2,-z_1)\mP_{12}\mE_{\ud}(-z_2,-z_1)
\nonumber\\
&&=\mF_{\ud}R_{23}K_2R_{32}\underbrace{R_{12}(z_1,-z_2)K_3}_{\textrm{commute}}R_{13}(z_1,-z_3)K_1R_{31}(-z_3,-z_1)R_{21}(-z_2,-z_1)\cdots\nonumber\\
&&\qquad\qquad \cdots \mP_{12}\mE_{\ud}(-z_2,-z_1)
\nonumber\\
&&=\mF_{\ud}\underbrace{R_{23}K_2R_{32}K_3}_{\textrm{Eq.}(\ref{BYBE})}R_{12}(z_1,-z_2)R_{13}(z_1,-z_3)K_1R_{31}(-z_3,-z_1)R_{21}(-z_2,-z_1)\cdots\nonumber\\
&&\qquad\qquad \cdots \mP_{12}\mE_{\ud}(-z_2,-z_1)
\nonumber\\
&&=\underbrace{\mF_{\ud}K_3}_{\textrm{commute}}R_{23}(z_2,-z_3)K_2
\underbrace{R_{32}(-z_3,-z_2)R_{12}(z_1,-z_2)R_{13}(z_1,-z_3)}_{\textrm{Eq.}(\ref{YBE})}K_1R_{31}(-z_3,-z_1)\cdots\nonumber\\
&&\qquad\qquad \cdots R_{21}(-z_2,-z_1)\mP_{12}\mE_{\ud}(-z_2,-z_1)
\nonumber\\
&&=K_3\mF_{\ud}R_{23}(z_2,-z_3)\underbrace{K_2R_{13}(z_1,-z_3)}_{\textrm{commute}}R_{12}(z_1,-z_2)\underbrace{R_{32}(-z_3,-z_2)K_1}_{\textrm{commute}}R_{31}(-z_3,-z_1)\cdots\nonumber\\
&&\qquad\qquad \cdots R_{21}(-z_2,-z_1)\mP_{12}\mE_{\ud}(-z_2,-z_1)
\nonumber\\
&&=K_3\mF_{\ud}R_{23}(z_2,-z_3)R_{13}(z_1,-z_3)K_2R_{12}(z_1,-z_2)K_1\cdots\nonumber\\
&&\qquad\qquad 
\cdots\underbrace{R_{32}(-z_3,-z_2)R_{31}(-z_3,-z_1)R_{21}(-z_2,-z_1)}_{\textrm{Eq.}(\ref{YBE})} \mP_{12}\mE_{\ud}(-z_2,-z_1)
\nonumber\\
&&=K_3\mF_{\ud}R_{23}(z_2,-z_3)R_{13}(z_1,-z_3)K_2R_{12}(z_1,-z_2)K_1\cdots\nonumber\\
&&\qquad\qquad \cdots \underbrace{R_{21}(-z_2,-z_1)R_{31}(-z_3,-z_1)R_{32}(-z_3,-z_2)\mP_{12}}_{\mP_{12}^2=1}\mE_{\ud}(-z_2,-z_1)
\nonumber\\
&&=K_3\mF_{\ud}R_{23}(z_2,-z_3)R_{13}(z_1,-z_3)K_2R_{12}(z_1,-z_2)K_1\cdots\nonumber\\
&&\qquad\qquad \cdots \mP_{12}\underbrace{R_{12}(-z_2,-z_1)}_{\textrm{Eq.}(\ref{REF})~\textrm{and}~\mP_{12}^2=1}R_{32}(-z_3,-z_1)R_{31}(-z_3,-z_2)\mE_{\ud}(-z_2,-z_1)
\nonumber\\
&&=K_3\mF_{\ud}R_{23}(z_2,-z_3)R_{13}(z_1,-z_3)\underbrace{K_2R_{12}(z_1,-z_2)K_1R_{21}(-z_2,-z_1)}_{\textrm{Eq.}(\ref{BYBE})}\mP_{12}\mE_{\ud}(-z_2,-z_1)\cdots\nonumber\\
&&\qquad\qquad \cdots \mF_{\ud}(-z_2,-z_1)R_{32}(-z_3,-z_1)R_{31}(-z_3,-z_2)\mE_{\ud}(-z_2,-z_1)
\nonumber\\
&&=K_3\mF_{\ud}R_{23}(z_2,-z_3)R_{13}(z_1,-z_3)\underbrace{R_{12}}_{\textrm{Eq.}(\ref{REF})}K_1R_{21}K_2\mP_{12}\mE_{\ud}(-z_2,-z_1)R_{3\ud}(-z_3|-z_2-z_1)
\nonumber\\
&&=K_3\mF_{\ud}\underbrace{R_{23}(z_2,-z_3)R_{13}(z_1,-z_3)R_{12}}_{\textrm{Eq.}(\ref{YBE})}\mE_{\ud}K_{\ud}(z_1,z_2)R_{3\ud}(-z_3|-z_2-z_1)
\nonumber\\
&&=K_3\underbrace{\mF_{\ud}R_{12}}_{\textrm{Eq.}(\ref{HF})}R_{13}(z_1,-z_3)R_{23}(z_2,-z_3)\mE_{\ud}K_{\ud}(z_1,z_2)R_{3\ud}(-z_3|-z_2-z_1)
\nonumber\\
&&=\underbrace{K_3\mH_{\ud}}_{\textrm{commute}}\mF_{\ud}R_{13}(z_1,-z_3)R_{23}(z_2,-z_3)\mE_{\ud}K_{\ud}(z_1,z_2)R_{3\ud}(-z_3|-z_2-z_1)
\nonumber\\
&&=\mH_{\ud}(z_1,z_2)K_3(z_3)R_{\ud3}(z_1,z_2|-z_3)K_{\ud}(z_1,z_2)R_{3\ud}(-z_3|-z_2-z_1)
\qquad \hfill\square
\ee
}

The proof of the complementary boundary fusion formula
(\ref{compKfused}) is similar to the one for the bulk
\cite{Beisert:2015msa}.  In particular, one first needs the identity
\be
\mF_{\ud} K_{1} R_{21}(z_{2},-z_{1}) K_{2} {\cal P}_{12} 
\bar\mE_{\langle \overline{12}\rangle}(-z_{2},-z_{1}) = 0\,,
\label{compfusid}
\ee
whose proof is as follows:
\be
\lefteqn{\underbrace{\mH_{\ud} \mF_{\ud}}_{\textrm{Eq.}(\ref{HF})} K_{1} R_{21} K_{2} {\cal 
P}_{12} \bar\mE_{\langle \overline{12}\rangle}(-z_{2},-z_{1})}\non \\
&&=\mF_{\ud} \underbrace{R_{12}  K_{1} R_{21} K_{2}}_{\textrm{Eq.}(\ref{BYBE})} {\cal 
P}_{12} \bar\mE_{\langle \overline{12}\rangle}(-z_{2},-z_{1}) \non\\
&&=\mF_{\ud} K_{2}  R_{12}(z_{1},-z_{2})  K_{1} \underbrace{R_{21}(-z_{2},-z_{1}) {\cal 
P}_{12}}_{{\cal 
P}_{12}^2=1} \bar\mE_{\langle \overline{12}\rangle}(-z_{2},-z_{1}) \non\\
&&=\mF_{\ud} K_{2}  R_{12}(z_{1},-z_{2})  K_{1} {\cal 
P}_{12} R_{12}(-z_{2},-z_{1}) \bar\mE_{\langle \overline{12}\rangle}(-z_{2},-z_{1}) = 0 \,.
\ee
In passing to the final equality, we used the fact 
$R(z_1,z_2)\bar\mE(z_1,z_2)=0$, which is a direct consequence of the 
decomposition (\ref{decomposition}) and the orthogonality relation $\mF(z_{1}, z_{2})\, \bar\mE(z_{1}, z_{2}) 
= 0$ (\ref{complopdef}).

It follows from (\ref{compfusid}) and the completeness relation (\ref{completeness}) that
\be
\lefteqn{\bar\mE_{\langle \overline{12}\rangle} 
\bar\mF_{\langle \overline{12}\rangle} K_{1} R_{21}(z_{2},-z_{1}) K_{2} {\cal P}_{12} 
\bar\mE_{\langle \overline{12}\rangle}(-z_{2},-z_{1})}\non\\
&&=\left(1-\mE_{\ud} \mF_{\ud}\right) K_{1} R_{21}(z_{2},-z_{1}) K_{2} {\cal P}_{12} 
\bar\mE_{\langle \overline{12}\rangle}(-z_{2},-z_{1})\non\\
&&=K_{1} R_{21}(z_{2},-z_{1}) K_{2} {\cal P}_{12} 
\bar\mE_{\langle \overline{12}\rangle}(-z_{2},-z_{1}) \,.
\ee
In other words, the projector $\bar\mE_{\langle \overline{12}\rangle} \bar\mF_{\langle \overline{12}\rangle}$ can be 
inserted or removed in front of 
\be
K_{1} R_{21}(z_{2},-z_{1}) K_{2} {\cal P}_{12} 
\bar\mE_{\langle \overline{12}\rangle}(-z_{2},-z_{1}) \non
\ee
as needed.  Armed with this fact, together with the corresponding 
bulk result \cite{Beisert:2015msa}
\be
\bar\mE_{\langle \overline{12}\rangle} \bar\mF_{\langle \overline{12}\rangle} 
R_{13} R_{23} \bar\mE_{\langle \overline{12}\rangle} = 
R_{13} R_{23} \bar\mE_{\langle \overline{12}\rangle}\,,
\ee
it is now a somewhat long but straightforward calculation to verify
that the complementary fused boundary K-matrix (\ref{compKfused})
obeys the fused boundary Yang-Baxter equation
\be
\lefteqn{\bar R_{\langle\overline{12}\rangle
3}(z_{1},z_{2}|z_{3})\, \bar K_{\langle \overline{12}\rangle}(z_{1}, z_{2})\, \bar R_{3 \langle \overline{12}\rangle}(z_{3}| 
-z_{2}, -z_{1})\, K_{3}(z_{3})} \non \\
&&= K_{3}(z_{3})\, \bar R_{\langle \overline{12}\rangle 3}(z_{1},z_{2}| -z_{3})\,
\bar K_{\langle \overline{12}\rangle}(z_{1}, z_{2})\,
\bar R_{3 \langle \overline{12}\rangle}(-z_{3}| -z_{2}, -z_{1}) \,,
\label{compfusedBYBE}
\ee
where the complementary fused R-matrices are given by (\ref{compRBA}) 
and (\ref{compRAB}).


\providecommand{\href}[2]{#2}\begingroup\raggedright\endgroup

\end{document}